# How Many Equations of Motion Describe a Moving Human?

Gabriele De Luca, Department for e-Governance and Administration, University for Continuing Education Krems, gabriele.deluca@donau-uni.ac.at (corresponding author)

Thomas J. Lampoltshammer, Department for e-Governance and Administration, University for Continuing Education Krems, thomas.lampoltshammer@donau-uni.ac.at

Johannes Scholz, Graz University of Technology, Institute of Geodesy, Research Group Geoinformation, johannes.scholz@tugraz.at

## Abstract

A human is a thing that moves in space. Like all things that move in space, we can in principle use differential equations to describe their motion as a set of functions that maps time to position (and velocity, acceleration, and so on). With inanimate objects, we can reliably predict their trajectories by using differential equations that account for up to the second-order time derivative of their position, as is commonly done in analytical mechanics. With animate objects, though, and with humans, in particular, we do not know the cardinality of the set of equations that define their trajectory. We may be tempted to think, for example, that by reason of their complexity in cognition or behaviour as compared to, say, a rock, then the motion of humans requires a more complex description than the one generally used to describe the motion of physical systems. In this paper, we examine a real-world dataset on human mobility and consider the information that is added by each (computed, but denoised) additional time derivative, and find the maximum order of derivatives of the position that, for that particular dataset, cannot be expressed as a linear transformation of the previous. In this manner, we identify the dimensionality of a minimal model that correctly describes the observed trajectories. We find that every higher-order derivative after the acceleration is linearly dependent upon one of the previous time-derivatives. This measure is robust against noise and the choice for differentiation techniques that we use to compute the time-derivatives numerically as a function of the measured position. This result imposes empirical constraints on the possible sets of differential equations that can be used to describe the kinematics of a moving human.

# 1. Introduction

If a kinematic system is a physical system that possesses some conserved quantities such as momentum or energy, it is then possible to derive its equations of motion analytically by using the Hamiltonian or Lagrangian formalism of classical mechanics. If the same system does not possess conserved quantities, then neither can be applied to describe the kinematic of the system (but see Sandler (2014) for extensions of the Lagrangian approaches to include dissipative systems such as living systems). And yet, some systems without conserved quantities present regularities in their trajectories; that is to say, the objects that comprise them may not move randomly and can thus not be described as simply undertaking Brownian motion. Humans that move in space are some such objects: because their motion and the selection of paths appears teleologically oriented to reaching some destinations (Golledge, 1995), and because the study of human mobility identified several factors that explain it (Golledge, 2003), it appears intuitively that it should be possible to describe such motion in terms of the mathematical language that is commonly used to describe the a motion of kinematic systems: the language of differential equations, by means of which the trajectory of a moving object can be modelled. Because we cannot however rely on the theoretical formalisms that are used to describe the motion of an inanimate system, we cannot aprioristically impose constraints on the shape that the equations of motion that describe a moving human should possess. In the Lagrangian formalism, for example, the equations of motion are derived by first setting up the Lagrangian of a system in terms of its potential and kinetic energy, and by then partial differentiating it with respect to the generalised coordinates and velocity, and by finally applying the Euler-Lagrange to extract the equations of motion. This, however, implies prior knowledge of the shape that the Lagrangian has, as a time-varying difference between the kinetic and potential energy of the system. In Hamiltonian formalism, similar reasoning is based upon the principle of conservation of energy, such that the total kinetic and potential energy of a system remains unvaried. And yet, these two approaches cannot possibly work out-of-the-box to describe human motion: humans, in fact, are not necessarily systems with conserved quantities; and even though their motion appears characterised by regularities, these regularities may not satisfy any of the symmetries that are commonly associated with the motion of physical systems with conserved quantities.

We would however appreciate it greatly if some Aristotle or some Newton came with a set of equations that describe the motion of a human as a function of time: this would let us, among other things, help solve the problem of modelling the trajectory of a human given the time series containing its position, without having to rely on deep learning as is instead done today (Chen, Li, Lu, & Zhou, 2021; Rossi, Paolanti, Pierdicca, & Frontoni, 2021). A legitimate question thus arises, in an attempt to find these equations of motion: how many these equations should be, and what would they look like? If we had this number, we could begin searching for them by means of regression over the observed or measured trajectories of the humans that roam around.

In this paper, we answer this question on the basis of the empirical trajectories measured about some humans, as retrieved from an openly-accessible dataset of human trajectories. In doing so, we explain that the dimensionality of a set of equations of motion for a moving human is finite and, for the particular dataset that we use as the empirical component of the research, we describe a method for identifying how many these equations are.

## 2. The Kinematic Modelling in Mechanics

The development of a measuring apparatus that observes the position of humans with high frequency and with high resolution call for research on the kinematics of human mobility. Some apparatus, such as the GPS contained in cheap smartphones, allow sampling of the position of a static object with an accuracy comprised within 1 to 4 centimetres of the ground truth (Uradziński & Bakuła, 2020). For a smartphone that measures position via a dual frequency GPS/Galileo system, the accuracy in kinematic motion is instead in the order of a metre (Elmezayen & El-Rabbany, 2019). Technologies alternative to the GNSS that allow tracking the position of a moving human also exist, and include the ultra-wide band ground radar (Chang, Wolf, & Burdick, 2010) or multimodal sensors that make use of lidar-radar fusion (Kwon, Hyun, Lee, Lee, & Son, 2017). Motion sensors attached to the human body can also be used for this purpose.

Knowledge of the instantaneous position of a human, knowledge of the instantaneous velocity and of some unknown additional number of time-derivatives of the position, determine the position of that same human at the next instant. In the real world, where the continuous sampling of a variable is not possible, it is sufficient to perform a frequent enough measurement so as to minimise the incidence that partial observation of the system has on the reliability of the statistical generalisations that are drawn from the measured samples (Abeliuk, Huang, Ferrara, & Lerman, 2020). As measuring apparatuses become increasingly more accurate, and sample at an increasingly higher resolution, this assumption will become increasingly more appropriate. In the simulated worlds, even today, this assumption is valid as it stands: if one had a set of equations that describe the time-evolution of the position of a human, and its associated time-derivatives, one could then integrate them numerically by using any arbitrarily small time interval and thus derive simulated trajectories that correspond to the ones described by the equations of motion.

However, we do not have these equations, no matter how much we would like them. Luckily, for inanimate objects, there came Newton who stated that the second order time derivative of the position of an object is proportional to some force that is applied on that object, such that:

$$\frac{d^2x}{dt^2}(t) = kF(t)$$

It can then be observed that by measuring the extent of the force with independent apparatus (e.g., a spring), and by measuring the weight of an object with another independent apparatus (e.g., a scale), this relationship is empirically observable if and only if the proportionality constant $k$ corresponds to the constant ratio between some fixed number $g$, common for all objects and independent of size or weight or speed, and the number $W$ corresponding to the weight of that object as measured on a graded scale. Under this condition, the proportionality constant corresponds to $k = \frac{g}{W} = \frac{1}{m}$, and the previous equation thus becomes the familiar:

$$\frac{d^2x}{dt^2}(t) = \frac{F(t)}{m}$$

If an analytical expression for $F(t)$ can be provided, such as "the force does not change with time and is thus constant", then $F(t) = $ Constant and the equation describes the motion of an object with uniform acceleration (e.g., in free fall). By then integrating twice with respect to time, and by solving for two independent boundary conditions (in this case, the position and the velocity), the two constants of integration can be determined. Then, this twice-integrated differential equation with the two constants of integration that correspond to the boundary conditions yields the equation that maps time to the position of an object, and allows to predict it infinitely in the future and in the past, insofar as the underlying dynamic of its motion remains unchanged.

## 3. The Kinematic Modelling of Human Motion

We don't however have a Newton who can explain to us what are the rules that the equations describing a moving human should follow; therefore, we are allowed to start reasoning from the empirical measurements concerning human trajectories first and then try to generalise them into some kind of expressions that would work regardless of the particular human or the particular trajectories that one observes. We would very much like to be able to do something similar to what is done in classical and analytical mechanics for human motion, in the sense that we would like to be able to express the trajectories followed by moving humans in terms of equations that associate time to position and its time derivatives. Certainly, there are empirical constraints that correspond to the shape that these equations can have: for example, we know that the speed of a human moving on foot is in the range $[0,11]\ ms^{-1}$, where the upper bound is slightly above the average speed of Usain Bolt during the competition that won him the Olympic medal (Eriksen, Kristiansen, Langangen, & Wehus, 2009). This implies that the module of the first-order time derivative of the position is always contained within that range, and this imposes constraints on the possible functions that can describe it; for example, the time derivative of the position cannot be a strictly increasing linear function of time. We also know that a human that is subject to acceleration or jerk higher than a certain threshold that varies according to their angle with that human's orientation, can cease to be a human very rapidly (Eiband, 1959). While this latter statement does not necessarily apply to the motion of humans on foot, it is nonetheless important to think that the system we study has some physical constraints that make its motion strongly non-random, even without considering the constraints that derive from the behavioural analysis conducted within social sciences, such as the observation that each human is found predominantly within one of two and only two locations at any given time, which we commonly refer to as "home" and "work" (or school, or equivalent) (Bagrow & Lin, 2012). But also, there are constraints on the equations that derive from the sake of our convenience: for example, we would like to have a system of equations that is as small as it possibly can (but not any smaller!), both in terms of the number of equations and its symbolic representation.

If, for example, we are describing the motion of a stationary human who is always found at position $x(t) = x_0$, then it would be sufficient to express the (degenerate) equation of motion for that human as its corresponding differential equation $\frac{dx}{dt} = 0$ with the given boundary condition that $x(t_0) = x_0$, and not for example as the system:

$$\frac{dx}{dt} = 0$$
$$\frac{d^2x}{dt^2} = 0$$

The second equation, while true, does not provide any additional information that is not already contained in the first. This consideration may seem trivial, but as we approach the problem of identification of the equations of motion for a human, we should keep in mind this particular example: the reason why a differential equation involving the second-order time derivative of the position does not matter, in this particular example, is that it can be expressed as a linear function of the first-order time derivative: $\frac{d^2x}{dt^2} = a\frac{dx}{dt} + b$, with $a = 1$ and $b = 0$. It will become apparent later why this particular method for expressing the relationship between acceleration and velocity is chosen. The acceleration thus contributes nothing, in this particular example, in terms of predictive power over the future position of that human, and should thus not be included in the set of equations that describe their motion.

If, however, the human was moving at a constant velocity $v_0$, it would then make sense to express its position as $x(t) = x_0 + v_0 t$ and its corresponding differential equation as:

$$\frac{dx}{dt} = v_0$$

Or, equivalently, to treat $x_0$ and $v_0$ as boundary conditions and express the system instead as:

$$\frac{d^2x}{dt^2} = 0$$

These last two expressions are equivalent. It would not make sense to include the third-order time derivative of the position, since it is unchanged and, notably, because it can be expressed as a linear function of the acceleration.

In other words, we would like to find a set of equations of motion that is the smallest possible set that provides enough predictive power into the trajectory of a moving human: if two such sets exist, we choose the one that, after appropriate substitutions of the various variables contained in it, can be maximally simplified and can thus contain the smallest number of differential equations. The question that we address in the next sections is: what is this number, in relation to a set of real-world human trajectories? And more generally, how can we find this number in absence of a theory of human motion that is analogous to the kinematic theories that are used for studying inanimate objects?

## 4. Data and methods

The source of empirical data that we use to answer this question is *MobilityModels* (Rhee et al., 2009), a dataset of 225 GPS trajectories that correspond to the motion of as many distinct individuals who volunteered for the purpose of recording their trajectories. The trajectories relate to five different locations: the university campuses of NCSU and KAIST, plus the urban area of New York, Disney World in Orlando, and the state fair in North Carolina. This dataset found previous usage in Chon, Shin, Talipov, and Cha (2012), where it was used to validate the applicability of Markov models; and also to validate the NextLocation models previously developed by Scellato, Musolesi, Mascolo, Latora, and Campbell (2011). In our case, however, consistent with the theoretical reasoning initiated in the previous section, we want to use this dataset in order to find constraints on the characteristics that the equations of motion describing the trajectories of all observed humans can have. The approach that we follow is grounded upon the bottom-up identification of rules for the dynamic evolution of non-linear systems, initially developed by Brunton, Proctor, and Kutz (2016), or the weaker version of SINDy, developed for the purpose of classifying individuals in a population (Messenger, Wheeler, Liu, & Bortz, 2022). Here we follow their underlying insight, that the identification of sparse and most simple rules can be done from measurements and from measurements alone if theories on the underlying dynamics are absent, but the specific implementation of the approach that we propose here is part of the innovative contribution of this research.

In each trajectory, the position of the humans is sampled with respect to a frame of reference that is stationary with respect to Earth. The position is measured along two orthogonal bases, and we do not have information as to the orientation of the frame of reference with respect to the geodetics of Earth, nor do we have information as to whether the same frame of reference is used in all trajectories that belong to the same region.

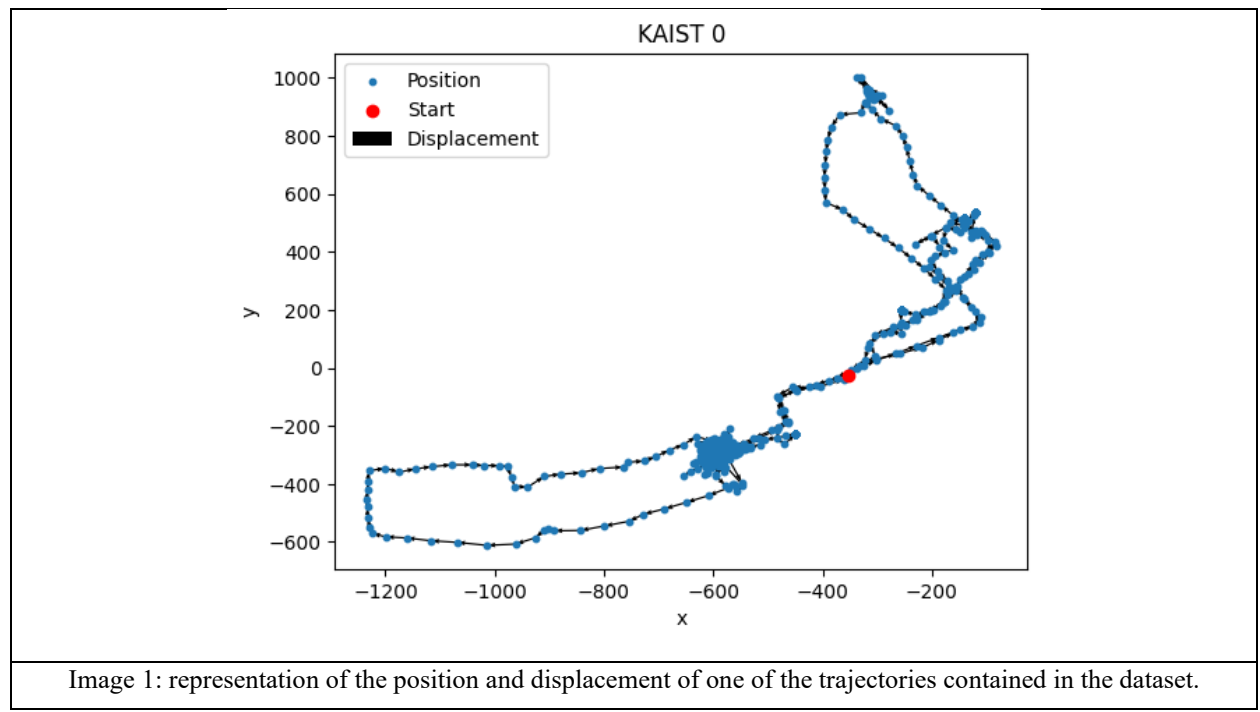

Image 1: representation of the position and displacement of one of the trajectories contained in the dataset.

The trajectories contained in this dataset are sampled at $\Delta t = 30\ s$, which is not particularly frequent; but for the purpose of this discussion, they will suffice. We are relying on a dataset collected by others and assume that the measurement of the position of humans at higher frequencies is possible: for this paper, we are not interested in the specific form for the equations of motion that we identify; but rather, in finding constraints that these equations must have in order to facilitate their identification and make them learnable. We then compute the velocities and the higher-order time derivatives by the smoothed finite-difference method (Farjadpour et al., 2006), which has the advantage over the finite-difference approximation of being more resistant to noise in the measurements. We assume that some system of differential equations such as the one shown below, corresponding to unknown functions of the position and its various time derivatives, represent the kinematic of this system:

$$\begin{cases} x(t) = c_{1,1} \times f_1(t) + 0 \times g_1(x) + c_{1,3} \times h_1(\frac{dx}{dt}) + c_{1,4} \times i_1(\frac{d^2x}{dt^2}) + \cdots \\ \frac{dx}{dt}(t) = c_{2,1} \times f_2(t) + c_{2,2} \times g_2(x) + 0 \times h_2(\frac{dx}{dt}) + c_{2,4} \times i_2(\frac{d^2x}{dt^2}) + \cdots \\ \frac{d^2x}{dt^2}(t) = c_{3,1} \times f_3(t) + c_{3,2} \times g_3(x) + c_{3,3} \times h_3(\frac{dx}{dt}) + 0 \times i_3(\frac{d^2x}{dt^2}) + \cdots \\ \vdots \\ \frac{d^nx}{dt^n}(t) = c_{n,1} \times f_n(t) + c_{n,2} \times g_n(x) + c_{n,3} \times h_n(\frac{dx}{dt}) + c_{n,4} \times i_n(\frac{d^2x}{dt^2}) + \cdots \end{cases}$$

If the problem is posed in this manner, and if the possible functions that correspond to the various functions of time and state variables are sparse in the domain of possible functions, then there are known techniques to learn these functions automatically on the basis of some sufficiently fine sets of measurements (Brunton et al., 2016). The applicability of the latter, however, depends upon the knowledge of the size of the set of equations that define the system; in this case, upon the prior knowledge of $n$, the order of the highest order time derivative that figures explicitly in the set of equations of motion.

We are particularly interested in identifying linear differential equations that allow expressing some higher-order time derivatives in terms of the lower-order ones; if they exist, and if they can consistently be found for all time derivatives higher than a certain order $n$, then this imposes $n$ as the dimensionality of the set of equations of motion that define the system. In other words, if it can be found that, for every $m \geq n$, there exists some $k \in \mathbb{N}_0$ such that $\frac{d^{m+k}x}{dt^{m+k}} = a\frac{d^mx}{dt^m} + b$, then each higher order derivatives of $x$ above the $n$-th can be expressed in terms of a lower one, and then the dimensionality of the minimal set of equations of motion that describe the system is $n$. The existence of such a relationship can be determined by means of pair-wise linear regression over the various time derivatives, which would result in identifying a coefficient $a$ in the equation above, close to $|a| = 1$, for higher-order derivatives that are linear in terms of some lower-order ones.

The hypothesis appears plausible via preliminary data analysis:

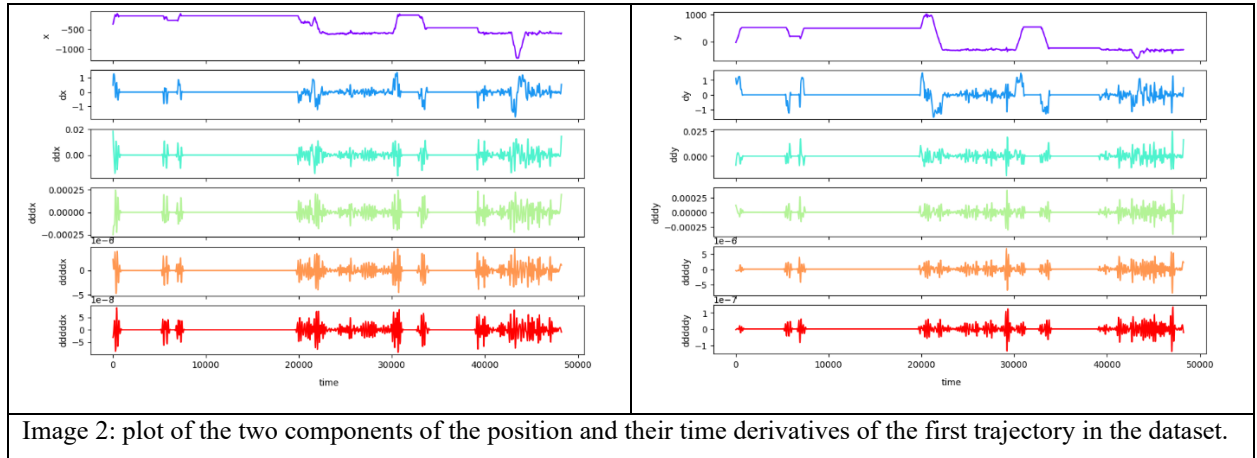

Image 2: plot of the two components of the position and their time derivatives of the first trajectory in the dataset.

We test this hypothesis and see if some value of $n$ can be identified for the dataset that we analyse. We study two versions of the trajectories and the associated time derivatives: a raw version, corresponding to the measurements contained therein, and a noise-filtered version that makes use of Kalman filtering, as commonly done when processing GPS signals (Moreno & Pigazo, 2009). This type of filtering assumes that the noise is Gaussian-distributed, which is a reasonable hypothesis in absence of evidence against it. The code by which the figures in the next section were generated is available on GitHub for the purpose of facilitating replication of the results.[1]

## 5. Results

The pair-wise Pearson correlation between the components of the position and each of its time derivatives was computed up to the 9-th time derivative. For the first trajectory, the correlation matrix shows a strong linear relationship between each time derivative and its corresponding acceleration (twice-differentiation):

---

[1] https://github.com/G-DL/Traces/blob/main/Traces3.ipynb

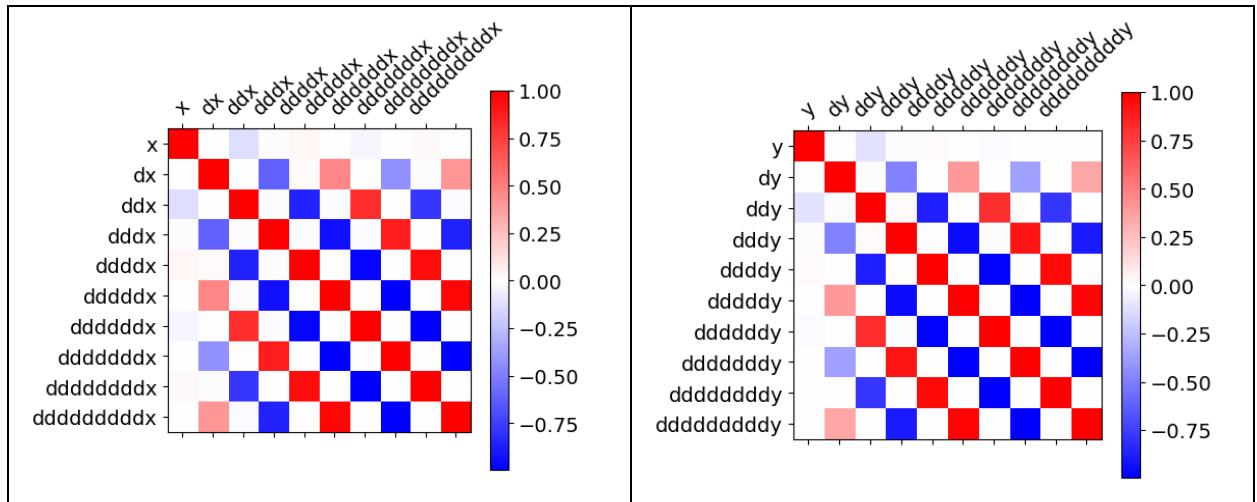

Image 3: Correlation matrix between the position and its time derivatives for the first, non-Kalman filtered trajectory in the dataset.

We computed the average correlation matrix for all the trajectories in the dataset, and extracted one for the unfiltered and one for the Kalman-filtered version of the trajectories. The figure below shows the two correlation matrices per component of the position:

| Component | $x(t)$ | $y(t)$ |
|---|---|---|
| Unfiltered | | |
| Filtered | | |

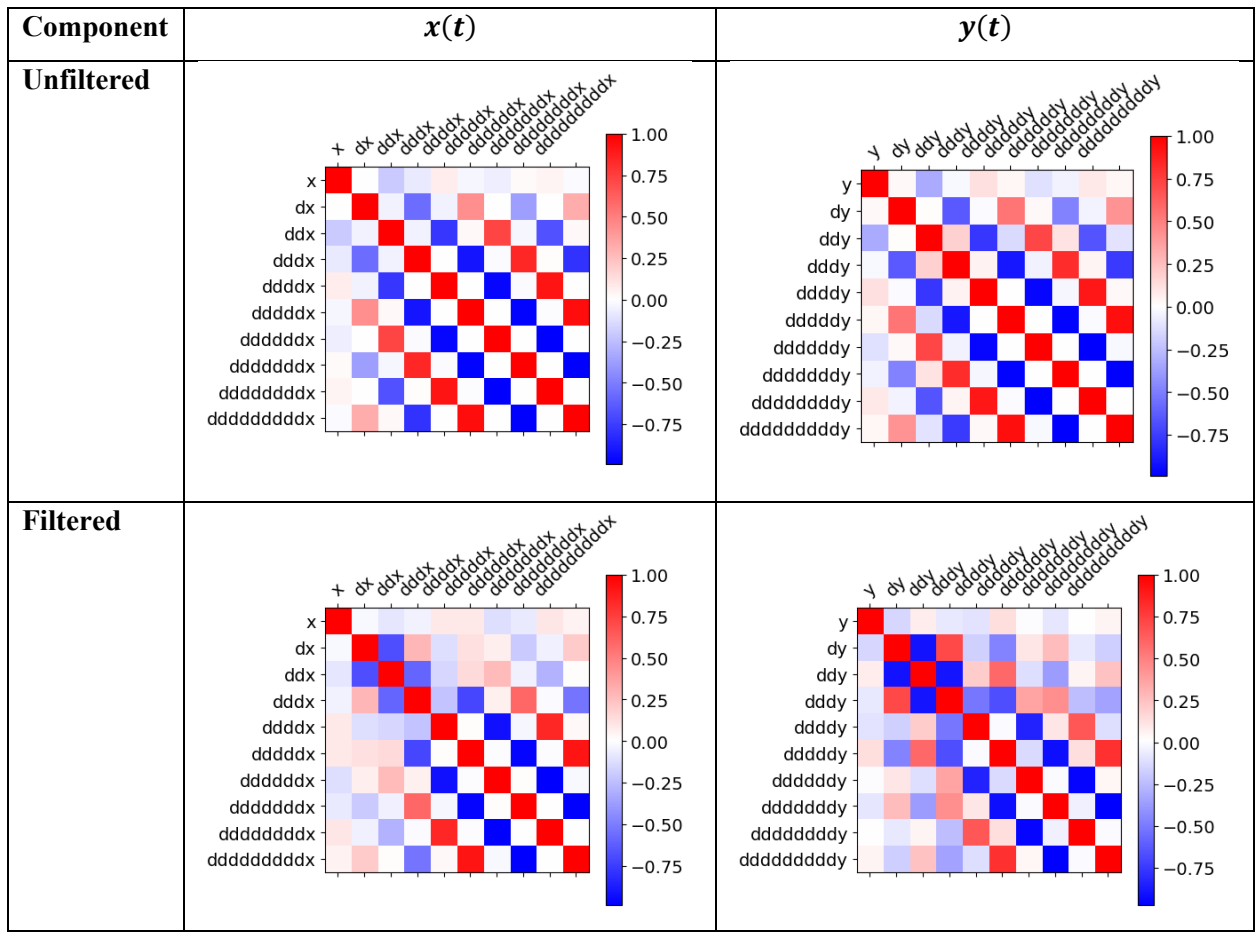

Image 4: average correlation matrices between each component of the position and its corresponding time-derivatives, as computed over the unfiltered and the Kalman-filtered trajectories of the dataset.

## 6. Discussion

For both the unfiltered and the filtered version of the trajectories, there is a strong linear relationship between some $n$-th order time derivative and its corresponding twice-derivative, which remains true for all tested derivatives from that order onward. For the unfiltered version of the dataset, this relationship is already identifiable starting from $\frac{d^4x}{dt^4} = a\frac{d^2x}{dt^2} + b$ onward; while for the filtered version of the dataset this relationship emerges between $\frac{d^2x}{dt^2} = a\frac{dx}{dt} + b$, between the velocity and the acceleration. Then, it follows the same pattern as for the unfiltered dataset and has the structure of $\frac{d^{m+2}x}{dt^{m+2}} = a\frac{d^m x}{dt^m} + b$ for $m >= 3$. This is an argument in favour of the existence of an unknown set of equations that describe the motion of humans in this trajectory, and in particular of the existence of one such set whose dimensionality is 3. This implies that, even though a human is not a deterministic system because its motion does not necessarily satisfy the conservation principles that describe them, there still are some unknown equations of motion that can describe the movement of humans as a function of time, and which result in the empirical trajectories being observed. The continuation of this study will consist of the expansion of the methodological approach proposed here to larger datasets comprising more trajectories and higher frequency sampling, and also in the concrete identification of the components of the equations of motion described here; for example, with the application of techniques for the sparse identification of non-linear functions that describe the evolution of a dynamical system.